\documentclass[12pt]{article}
\usepackage{amsfonts}
\usepackage{latexsym}
\usepackage{amssymb}
\date{}
\begin{document}

\title{Entropy Identity and Material-Independent\\Equilibrium
  Conditions\\in Relativistic Thermodynamics}
\author{W. Muschik\quad and\quad H.-H. v. Borzeszkowski
\footnote{Corresponding author: borzeszk@itp.physik.tu-berlin.de}\\
%H.-H. v. Borzeszkowski\hspace{.2cm} and\hspace{.2cm} H. Herrmann\\
Institut f\"ur Theoretische Physik\\
Technische Universit\"at Berlin\\
Hardenbergstr. 36\\D-10623 BERLIN,  Germany
}
\maketitle

            \newcommand{\be}{\begin{equation}}
            \newcommand{\beg}[1]{\begin{equation}\label{#1}}
            \newcommand{\ee}{\end{equation}\normalsize}
            \newcommand{\bee}[1]{\begin{equation}\label{#1}}
            \newcommand{\bey}{\begin{eqnarray}}
            \newcommand{\byy}[1]{\begin{eqnarray}\label{#1}}
            \newcommand{\eey}{\end{eqnarray}\normalsize}
            \newcommand{\beo}{\begin{eqnarray}\normalsize}
            \newcommand{\R}[1]{(\ref{#1})}
            \newcommand{\C}[1]{\cite{#1}}

            \newcommand{\mvec}[1]{\mbox{\boldmath{$#1$}}}
            \newcommand{\x}{(\!\mvec{x}, t)}
            \newcommand{\m}{\mvec{m}}
            \newcommand{\F}{{\cal F}}
            \newcommand{\n}{\mvec{n}}
            \newcommand{\argm}{(\m ,\mvec{x}, t)}
            \newcommand{\argn}{(\n ,\mvec{x}, t)}
            \newcommand{\T}[1]{\widetilde{#1}}
            \newcommand{\U}[1]{\underline{#1}}
            \newcommand{\X}{\!\mvec{X} (\cdot)}
            \newcommand{\cd}{(\cdot)}
            \newcommand{\Q}{\mbox{\bf Q}}
            \newcommand{\p}{\partial_t}
            \newcommand{\z}{\!\mvec{z}}
            \newcommand{\bu}{\!\mvec{u}}
            \newcommand{\rr}{\!\mvec{r}}
            \newcommand{\w}{\!\mvec{w}}
            \newcommand{\g}{\!\mvec{g}}
            \newcommand{\D}{I\!\!D}
            \newcommand{\se}[1]{_{\mvec{;}#1}}
            \newcommand{\sek}[1]{_{\mvec{;}#1]}}            
            \newcommand{\seb}[1]{_{\mvec{;}#1)}}            
            \newcommand{\ko}[1]{_{\mvec{,}#1}}
            \newcommand{\td}{{^{\bullet}}}
            \newcommand{\eq}{{_{eq}}}
            \newcommand{\eqo}{{^{eq}}}
            \newcommand{\f}{\varphi}
            \newcommand{\seq}{\stackrel{_\bullet}{=}}
\newcommand{\Section}[1]{\section{\mbox{}\hspace{-.6cm}.\hspace{.4cm}#1}}
\newcommand{\Subsection}[1]{\subsection{\mbox{}\hspace{-.6cm}.\hspace{.4cm}
\em #1}}
%eigene makros
\newcommand{\const}{\textit{const.}}
\newcommand{\vect}[1]{\underline{\ensuremath{#1}}}  %Vektoren
\newcommand{\abl}[2]{\ensuremath{\frac{\partial #1}{\partial #2}}}
%partielle Ableitung

\noindent
{\bf Abstract}
On the basis of the balance equations for energy-momentum, spin, particle and 
entropy density, an approach is considered which represents a comparatively 
general framework for special- and general-relativistic continuum 
thermodynamics. In the first part of the paper, a general entropy density 
4-vector, containing particle, energy-momentum, and spin density contributions,
is introduced which makes it possible, firstly, to judge special assumptions 
for 
the entropy density 4-vector made by other authors with respect to their 
generality and validity and, secondly, to determine entropy supply and 
entropy production. Using this entropy density 4-vector, in the second part, 
material-independent equilibrium conditions are discussed. While in literature,
at least if one works in the theory of irreversible thermodynamics assuming 
a Riemann space-time structure, generally thermodynamic equilibrium is 
determined by introducing a variety of conditions by hand, the present 
approach proceeds as follows: For a comparatively wide class of space-time 
geometries the necessary equilibrium conditions of vanishing entropy supply 
and entropy production are exploited and, afterwards, supplementary conditions 
are assumed which are motivated by the requirement that thermodynamic 
equilibrium quantities have to be determined uniquely.

\Section{Introduction}

Relativistic thermodynamics for an 1-component material starts out with 
the balance equations of the particle flux 4-vector $N^k$, the
energy-momentum tensor $T^{ik}$ and the spin tensor $S_{ji}^{\cdot\cdot k}$
\bee{1}
N^k_{\cdot\mvec{;}k}\ =\ 0,\qquad 
T^{ki}_{\cdot\cdot\mvec{;}k}\ =\ G^i + K^i ,\qquad
S_{ji\mvec{;}k}^{\cdot\cdot k}\ =\ H_{[ji]} + L_{[ji]}\ 
\footnote{Square brackets are also used to emphasize that
the tensor is antisymmetric, $L_{(ji)}=0$, especially for $H_{ji}$ and 
$L_{ji}$} 
\ee
and with the balance of the 4-entropy $S^k$
\bee{2}
S^k_{\cdot\mvec{;}k}\ =\ \f +\sigma,\qquad\sigma \geq 0.
\ee
As usual, the semicolon ''$_{\mvec{;}}$'' denotes the covariant derivative,
$T^{ik}$ is the energy momentum-tensor of a material which is
not necessarily symmetric with vanishing covariant derivative, the
spin tensor
$S_{ji}^{\cdot\cdot k}$ is skew-symmetric in the lower indices, and 
$\f$ and $\sigma$ are the entropy supply and the entropy production, 
respectively. The force $K^i$ and the angular momentum 
$L_{[ji]}$ are the external sources of the energy momentum tensor
and of the spin tensor. 
\vspace{.3cm}\newline%%%%%%%%%%%%%%%%%%%%%%%%%%%%%%%%%%%%%%
The supply terms $\f,\ K^i$, and $L_{[ji]}$ are
substitutes for cases in which the energy-momentum tensor and/or the
spin tensor do not include all fields, so that additional fields come 
into account by external sources. %%%EEEEEEEEEEEEEEEEEEE 
Of course, in a field theory describing systems completely by the equations 
of the fundamental fields, external sources do not occur. If one is forced 
to introduce supply terms, this shows that the theory is not 
field-theoretically complete. To complete it, one has to describe the supply 
terms by additional fundamental fields in such a way that they can be 
absorbed by the other expressions in the balances \R{1} \C{4,HHBCH}.
%%%EEEEEEEEE 
%Often, these sources can be represented 
%by divergence terms and therefore can be included in an effective 
%energy-momentum tensor or an effective spin tensor on the left-hand sides of
%\R{1}$_2$ and \R{1}$_3$. 
From a thermodynamical
point of view, this procedure to include the supplies into effective
tensors from the very beginning, is disadvantageous, as we will see below.
%%%EEEEEEEEEEEEEEEEEE
%Especially, in the case of Lagrangian-supported theories, supply terms 
%signals the fact that the Lagrangian does not completely describe the 
%physical system under consideration. 
Here, we imply supply terms for the following 
reason: %\newline 
Often one considers a situation in which an approximate or a phenomenological 
description is sufficient, and one does not need a complete 
description of the system. For example, one need not to imply Maxwell's 
equations (or the corresponding Lagrangian), if one only intends to regard 
the influence of a given external electromagnetic field on a charged fluid. 
This influence can be regarded by assuming that $K^i$ is given a by the 
Lorentz (volume) force.   
\vspace{.3cm}\newline%%%%%%%%%%%%%%%%%%%%%%%%%%%%%%%%%%%%%%
The $G^i$ and $H_{[ij]}$ are internal source terms caused by the
choice of a special space-time and by the spin-momentum-energy
coupling (SMEC). 
For instance in Einstein-Cartan
geometry, the $G^i$ and $H_{[ij]}$ are caused by the torsion and
depend as coupling terms on the energy-momemtum and on the spin tensor.   
We call a theory for which the $G^i$ and $H_{[ij]}$ vanish identically
SMEC-free. 
%Thus, the General Relativity Theory (GRT) is SMEC-free, if the
%spin tensor is zero. Taking the spin into consideration, the SMEC-term
%in GRT is the asymmetric part of the energy-momentum tensor: 
%$H_{[ik]}= L_{[ik]}$. The Einstein-Cartan theory is not SMEC-free. Here, the
%SMEC-terms are functions of the torsion and the energy-momentum tensor.
\vspace{.3cm}\newline%%%%%%%%%%%%%%%%%%%%%%%%%%%%%%%%%%%%%%
In contrast to Special Relativity Theory (SRT) and Einstein-Cartan 
Theory (ECT), General Relativity Theory (GRT), makes 
no general statements on the structure of spin and spin balances, except 
for that here does not occur a spin tensor as explicit source of gravity. 
The spin of the matter source has only an implicit influence on the 
gravitational field insofar, as the source term in Einstein's equations
(the symmetric metrical energy-momentum tensor), differs for different 
kinds of spinorial matter. 
%Accordingly, one is led to Einstein's equations 
%with different metrical energy-momentum tensors coupled to field equations 
%of different spinorial matter (e.g., of spin-1-field and spin-1/2-field 
%matter). 
In some cases, where the total set of equations consists of 
Einstein's equations coupled to field equations of phenomenological 
matter, one can derive from this set, beside the energy-momentum balance, 
also a spin balance. For a Weyssenhoff fluid, particularly follows beside   
$K^i = 0$ the SMEC-term $H_{[ik]}= T_{[ik]}$ \C{OBPI}.
\vspace{.3cm}\newline%%%%%%%%%%%%%%%%%%%%%%%%%%%%%%%%%%%%%%%%%%%%
The non-negative entropy production $\sigma$ in \R{2} represents the
strong formulation of the
Second Law of thermodynamics in field theories. The inequality
\bee{aa2}
S^k_{\cdot\mvec{;}k} - \f\ \geq\ 0
\ee
is called the dissipation inequality.
\vspace{.3cm}\newline%%%%%%%%%%%%%%%%%%%%%%%%%%%%%%%%%%%
The relations \R{1} and \R{2} are the relativistic generalization of the 
balance equations of non-relativistic continuum thermodynamics. In their 
special-relativistic version -- with vanishing spin tensor, 
vanishing supply terms and vanishing SMEC-terms --
they were introduced by Eckart \C{EC} and Kluitenberg \C{KL}.
The 
quantities appearing in \R{1} and \R{2} are tensors with respect to Lorentz 
transformations, and the derivatives denoted by the semicolon have to be read 
as partial derivatives, if one refers to inertial systems, and as covariant 
derivatives (with the Christoffel symbols as components of the Levi-Civita 
connection), if non-inertial reference systems are considered.
\vspace{.3cm}\newline%%%%%%%%%%%%%%%%%%%%%%%%%%%%%%%%%%%%%%%%%%%%%%%%
Assuming that dynamics takes place in a curved space-time, the equations 
\R{1} and \R{2} describing this dynamics have to be interpreted as generally 
covariant tensor equations in this chosen space-time. This means, that
the basic quantities in \R{1} and \R{2} 
must be considered as tensors with respect to arbitrary systems of reference 
(observers) and that the semicolon derivative is the covariant derivative 
defined by the space-time geometry belonging to the theory of gravitation 
under consideration.
\vspace{.3cm}\newline%%%%%%%%%%%%%%%%%%%%%%%%%%%%%%%%%%%%%%%%%%%%%%%
In case of GRT, a Riemann space-time is assumed
whose covariant derivative is given by the Levi-Civita connection. (The 
above-mentioned Minkowski space-time of special relativity theory is a Riemann 
space-time with vanishing curvature, i.e., the case for which gravitation 
curving the space-time is neglected.) In generalizations of GRT, instead of 
the Riemann space-time defined by the metric as a primary quantity, 
geometrically
more rich geometries are considered. For instance, one can introduce 
space-times that are characterized by additional quantities like torsion and 
non-metricity which are independent of the metric \C{3}. In these
space-times, non-vanishing SMEC-terms $G^i$ and $H_{[ij]}$ appear
which are describing the coupling of the torsion to spin and energy-momentum.
\vspace{.3cm}\newline%%%%%%%%%%%%%%%%%%%%%%%%%%%%%%%%%%%%%%%%%%%%%
As long as one does not consider the gravitational field equations specifying 
the space-time, but only a given curved space-time of one of the above 
mentioned types is assumed, one can work with the general framework given by 
\R{1} and \R{2}. However, the situation changes drastically, if completing 
the theory 
by incorporating the gravitational equations. In GRT, the gravitational 
field and thus the curvature of the Riemann space-time is determined by 
Einstein's equations
\byy{2a}
R_{ik}-\frac{1}{2}g_{ik}R &=& -\kappa \hat{T}_{ik},
\eey
and therefore one has to assume a symmetric energy-momentum tensor 
$\hat{T}_{ik}$ and vanishing $K^i$ ($G^i$ and $H_{[ik]}$ vanish in GRT, 
if there is no spin. In this case GRT is SMEC-free)
\byy{a2}
\hat{T}^{ik}\ =\ \hat{T}^{ki},\qquad\hat{T}^{ki}_{\cdot\cdot\mvec{;}k}\ 
=\ 0.\vspace{.3cm}
\eey
The condition \R{a2}$_1$ expresses the fact that in GRT way is given for 
matter of non-vanishing spin, but that the spin of matter is only insofar 
a source of gravitation, as it can be reflected by terms contained in the 
symmetric energy-momentum tensor $\hat{T}^{ik}$. If one considers matter 
with spin 
having a non-symmetric energy-momentum tensor of matter, this matter can be 
incorporated into GRT by symmetrizing the energy-momentum tensor, such that 
it satisfies \R{a2}. As to condition \R{a2}$_2$, it requires in \R{1}$_2$ a 
supply term $K^i$, of the form  $K^i =\Theta^{ik}_{\cdot\cdot\mvec{;}k}$, 
such that it 
can be rewritten into a divergence of a second-rank tensor $\hat{T}^{ik} - 
\Theta^{ik}$. In some cases it was shown that, starting out with conditions
\R{1}, \R{a2} can also be established (see e.g. \C{4}).
\vspace{.3cm}\newline%%%%%%%%%%%%%%%%%%%%%%%%%%%%%%%%%%%%%%%%%%%%%%%
For generalizations of GRT like Einstein-Cartan theories (see
e.g. \C{5,6,7}), due 
to the changed geometry, one finds generalized Bianchi identities and 
gravitational equations modifying Einstein equations \R{2a}. Also in these
cases, one has to arrange that these new identities and equations are 
compatible with the balance equations. Thus, again, one obtains restrictions 
to the balances or to the material field equations.
\vspace{.3cm}\newline%%%%%%%%%%%%%%%%%%%%%%%%%%%%%%%%%%%%%%%%%%%%%%%
In this paper, we do not assume a special theory of gravitation, but we discuss
the balance equations \R{1} and \R{2} for the general case of a curved 
space-time with a given background specifying metric and connection. Attention
is concentrated on the analysis of the dissipation inequality \R{2}
from the point of view that it has to be satisfied by any ansatz for the 
entropy vector. The results are generally valid in SRT and have to be specified
by additional conditions, especially for $G^i$ and $H_{[ij]}$, 
in the case of non-SMEC-free relativistic gravitational theories.
\vspace{.3cm}\newline%%%%%%%%%%%%%%%%%%%%%%%%%%%%%%%%%%%%%%%%%%%%%%%
For solving the system of differential
equations \R{1} and \R{2} in different chosen geometries according to \R{2a} 
or according to other gravitational theories, constitutive equations are
needed, because the balances and field equations are valid for
arbitrary, for the present unspecified materials.
Here $N^k , T^{ik}, S^k$ and $S_{ik}^{\cdot\cdot l}$ are constitutive 
mappings defined on a large state space (no after-effects) \C{MUPAEH96}
\bee{3}
{\bf z} = (g_{ik},{\cal T}_{ik}^{\cdot\cdot l},n, e, s_{ik}
\Xi_k , u^k , .......),
\ee
which may contain the geometrical fields, such as the metric $g_{ik}$, 
the torsion ${\cal T}_{ik}^{\cdot\cdot l}$, and 
the wanted basic fields $(n, e, s_{ik}, \Xi_k , u^k )$ (particle number
density, energy density, spin density, spin density vector
and an other for the present arbitrary time-like vector field $u^k$) 
and beyond them
other fields ..... which depend on the considered material and which are of no
interest here, because we are looking for {\em material-independent}
properties. Consequently, a special constitutive equation will not
appear in this paper. 
\vspace{.3cm}\newline%%%%%%%%%%%%%%%%%%%%%%%%%%%%%%%%%%%%%%%%%%%%%%%
In relativistic irreversible thermodynamics, stable thermodynamical 
equilibria are characterized by the fact, that the temperature
4-vector is a Killing vector \C{NEU80}. But this is only true, if $T^{ik}$
is symmetric and if the space-time is SMEC-free, properties which are 
not valid in general and which are not
presupposed here. Therefore the question arises and will be answered:
Are there equilibrium conditions independently of constitutive
properties in the framework of a general gravitation theory?
\vspace{.3cm}\newline%%%%%%%%%%%%%%%%%%%%%%%%%%%%%%%%%%%%%%%%%%%%%% 
The paper is organized as follows: starting out with the 3-1-split of
the quantities appearing in \R{1} and \R{2}, 
%\C{NEU80,10} 
we derive 
an identity for the entropy 4-vector \C{DPG90} in Sect.\ref{EI}. Using 
Eckart's interpretation
of the time-like vector $u^k$ as 4-velocity of the material \C{EC} in
Sect.\ref{LL}, we are formulating material-independent equilibrium 
conditions in Sect.\ref{EC}. Proofs can be found in the appendices.

\Section{3-1-Split}

The normalized time-like vector field $u^k$ included in the state space \R{3} 
(signature of the metric is --2) 
\bee{4}
u^k u_k \ =\ a^2 \ >\ 0 \quad\longrightarrow\quad
u^k u_{k;m}\ =\ 0
\ee
is first of all arbitrary and can therefore be chosen in
different ways. Here, it is introduced for spitting the quantities into
their parts parallel and perpendicular to $u^k$. This split allows for
a special interpretation later on. By introducing the projector
belonging to \R{4}
\bee{5}
h^k_l \ :=\ g ^k_l \ -\ \frac{1}{a^2}u^k u_l ,
\ee
we obtain as shown in appendix 1
\byy{6}
N^k \ =\ \frac{1}{a^2}nu^k \ +\ n^k ,\qquad S^k \ =\ \frac{1}{a^2}su^k
\ +\ s^k ,\\ \label{7}
T^{ik}\ =\ \frac{1}{a^4}eu^i u^k \ +\ \frac{1}{a^2}u^i p^k \ +\
\frac{1}{a^2}q^i u^k \ +\ t^{ik},\\ \label{7a}
S_{ik}^{\cdot\cdot l}\ =\ \left(\frac{1}{a^2}s_{ik}+\frac{1}{a^4}
u_{[i}\Xi_{k]}\right)u^l + s_{ik}^{\cdot\cdot l} +
\frac{1}{a^2}u_{[i}\Xi_{k]}^{\cdot l}.
\eey
Here the following abreviations are introduced
\byy{8}
n\ :=\ N^k u_k ,\quad n^k & :=& h^k_l N^l ,\hspace{3cm}\\  \label{9}
s\ :=\ S^k u_k ,\quad s^k  &:=& h^k_l S^l ,\hspace{3cm}\\  \label{10}
e := u_l u_m T^{lm} , \ p^k  := h^k_l u_m T^{ml},\quad
q^k  &:=& h^k_l u_m T^{lm},\ t^{ik} := h^i_l h^k_m T^{lm}\\ \label{10a}
s_{ik} := S_{ab}^{\cdot\cdot c}h^a_i h^b_k u_c ,\quad \Xi_k &:=&
2S_{ab}^{\cdot\cdot c}u_c u^a h^b_k ,\\ \label{10b}
s_{ik}^{\cdot\cdot l} := S_{ab}^{\cdot\cdot c}h^a_i h^b_k h^l_c ,\quad
\Xi_k^{\cdot l} &:=& 2S_{ab}^{\cdot\cdot c}h_c^l u^a h^b_k .
\eey
The physical interpretation of the quantities in \R{8} to \R{10b}
remains uncertain, as long as the the time-like vector field $u^k$ is not
interpreted. Later on, the quantities in \R{10a} and \R{10b} are recognized
as follows \C{HERMU04}: $s_{ik}$ is the spin density, 
$s_{ik}^{\cdot\cdot l}$
the couple stress, $\Xi_k$ the spin density vector and $\Xi_k^l$ is the
spin stress. These quantities are not independent of each other, but they are
coupled by the spin axioms \C{HERMU04} which we will use later.
Independently of any interpretation, the 4-entropy satisfies 
an identity which is derived in the next section.

\Section{The Entropy Identity\label{EI}}

In literature, one finds different approaches to a special- and 
general-relativistic conception of entropy. Most of them is in common 
that entropy is described by a 4-vector, but there are proposed different 
expressions for it (which generally do not incorporate spin terms). For 
instance, in \C{11} is generalized the non-relativistic expression of the 
internal energy $U$ for constant temperature and composition
\bee{8a}
U\ =\ TS - \mu n
\ee
($T$ = rest temperature, $S$ = entropy,
$\mu$ = chemical potential) which results by 
differentiation in classical thermodynamics of discrete systems
together with the Gibbs equation in the Gibbs-Duhem equation. This yields an
entropy 4-vector
\bee{8b}
S^k \ =\ \mu N^k + \frac{u_m}{T}T^{km} + p \frac{u^k}{T}\ =\
\mu N^k + \frac{u_m}{T}\left[T^{km}+ p\delta ^{km} \right]
\ee
($p$ = pressure).
\vspace{.3cm}\newline%%%%%%%%%%%%%%%%%%%%%%%%%%%%%%%%%%%%%%%
Other authors make an ansatz for the entropy vector such that its covariant 
divergence becomes a relativistic generalization of the Carnot-Clausius 
relation,
\bee{8c}
d_e S\ =\ \frac{\delta Q}{T}.
\ee
Here ``$d_e$'' denotes a change caused by an external supply (see
e.g. \C{NEU80}).
Accordingly, they assume\footnote{For the present, a vector $O_m$ is 
introduced which later on is identified to be equal to $u_m /T$.}
\bee{8d}
S^k \ =\ \mu N^k + \frac{u_m}{T}T^{km}.\vspace{.3cm}
\ee%%%%%%%%%%%%%%%%%%%%%%%%%%%%%%%%%%%%%%%%%%%%%%%
The procedure in this paper is quite different: We do not make ansatzes of the
entropy vector $S^k$, but we start out with an identity which runs as follows:
\vspace{.3cm}\newline$\Diamond$
Independently of the special interpretation of the time-like vector
field $u^k$, the following identity for the 4-entropy is valid:
\bee{13}
S^k \ \equiv\ (s^k - \lambda q^k - \mu n^k - \Lambda^m \Xi_m^{\cdot k}) 
+ (\mu N^k + \xi _l T^{kl} + \zeta^{nm}S_{nm}^{\cdot\cdot k}),
\ee
with the following abbreviations:
\byy{14}
\lambda \mbox{ arbitrary scalar},\ \Lambda^k \mbox{ arbitrary tensor
field of 1st order},\hspace{1.7cm}\\ \label{14a}
\mu\ :=\ \frac{1}{n}(s - \lambda e - \Lambda^m \Xi_m),\quad 
\xi _l \ :=\ \lambda u_l ,\quad \zeta^{nm} := 2u^n \Lambda^p h_p^m .
\hspace{1cm}\Diamond
\eey
$\Box$ The proof is easy: Starting out with the relations \R{6}
\bee{14a1}
s^k \ =\ S^k - \frac{s}{n}\left( N^k - n^k \right),
\ee
we obtain from \R{10}$_3$, \R{5}, \R{10}$_1$ and \R{6}$_1$
\bee{11}
q^k \ =\ u_m T^{km} - \frac{1}{a^2}eu^k \ =\ u_m T^{km} - \frac{e}{n}
\left( N^k - n^k \right). 
\ee
From \R{10b}$_2$ follows by use of \R{5}, \R{10a}$_2$ and \R{6}$_1$
\bey\nonumber
\Xi_m^{\cdot k} &=& 2u^p h_m^q S_{pq}^{\cdot\cdot k} - \frac{2}{a^2}
u^p h_m^q S_{pq}^{\cdot\cdot r}u^k u_r \ =\\ \label{11a}
&=& 2u^p h_m^q S_{pq}^{\cdot\cdot k} - \frac{1}{n}\Xi_m 
\left( N^k - n^k \right).
\eey
Summing up the last three equations multiplied with $\lambda$ and
$\Lambda^m$, we obtain
\bey\nonumber
&&\hspace{-1cm}s^k - \lambda q^k - \Lambda^m \Xi_m^{\cdot k}\ =\\ \label{11b} 
&& \hspace{-1cm}=\ S^k - \lambda u_m T^{km} - 2\Lambda^m u^p h_m^q 
S_{pq}^{\cdot\cdot k} + \frac{1}{n}\left(-s + \lambda e + 
\Lambda^m \Xi_m\right)\left( N^k - n^k \right)
\eey
which is identical to \R{13}.\hfill$\Box$
\vspace{.3cm}\newline
Consequently, the identity \R{13} is valid for arbitrary $\lambda$ and 
$\Lambda^m$ and for all time-like vector fields $u^k$. 
\vspace{.3cm}\newline%%%%%%%%%%%%%%%%%%%%%%%%%%%%%%%%%%%%%%%%%
The ad-hoc chosen entropy vector \R{8d} is in accordance with the 
identity \R{13} by setting 
\bee{b11}
\Lambda^m \ :=\ 0,\quad \zeta^{nm}\ :=\ 0,\quad \lambda\ :=\
\frac{1}{T},\quad s^k \ :=\ \lambda q^k + \mu n^k .
\ee
But it is not quite clear, if \R{8d} represents the most general ansatz 
also without spin,
since the identity \R{13} allows for adding a space-like vector,
the first bracket in \R{13}.
To clarify this question and for incorporating spin, we do not start out with
a specific ansatz for the entropy vector, but with the identity \R{13} in 
Sect.\ref{EB}.
\vspace{.3cm}\newline%%%%%%%%%%%%%%%%%%%%%%%%%%%%%%%%%%%%%%%%%
In contrast to the expression \R{8d} for the entropy 4-vector, \R{8b} is
not in accordance with the identity \R{13}. If the chemical potential
$\mu$ and the energy-momentum tensor $T^{km}$ in \R{8b} are the same
quantities as in \R{13}, we obtain for the spin-free case by comparing 
\R{8b} with \R{13} the false equation
\bee{c11}
p\frac{u^k}{T}\ \stackrel{f!}{=}\ s^k - \lambda q^k - \mu n^k .
\ee
Consequently, $\mu$ and $T^{km}$ in \R{8b} are different from those in
\R{13}, or \R{8b} is wrong. 
\vspace{.3cm}\newline%%%%%%%%%%%%%%%%%%%%%%%%%%%%%%%%%%%%%%%%%
Without any restriction of generality, from \R{13}, \R{14a}, \R{10a}$_2$ 
and \R{10b}$_2$ follows, that $\Lambda^m$ can be chosen orthogonal to $u^m$
\bee{11c}
\Lambda^m \ \doteq\ \Lambda^p h_p^m. 
\ee
Later on, this choice makes an interpretation of $\Lambda^m$ more easy.
\vspace{.3cm}\newline
In the next section, we will identify the time-like $u^k$ field, thus
resulting in an interpretation of the quantities \R{8} to \R{10b}.

\Section{Eckart and Landau-Lifshitz Interpretation\label{LL}}

Two different interpretations of the $u^k$ can be found in literature: 
the first one is due to Landau-Lifshitz \C{12}, the second one due to
Eckart \C{EC}. 
\vspace{.3cm}\newline
Landau-Lifshitz choose $u^k$ as an eigenvector of the energy-momentum
tensor
\bee{17}
u_m T^{km}\ =\ \frac{e}{a^2}u^k .
\ee
By \R{11}, this choice results in
\bee{18}
q^k \ \equiv\ 0 .
\ee
That means, this choice fixes 3 of the 16 free components of the 
energy-momentum tensor. Because this tensor represents a constitutive
mapping, \R{17} is introducing a special constitutive property.
Because we are looking for material independent statements, we do
not accept the Landau-Lifshitz choice \R{17} of $u^k$.
\vspace{.3cm}\newline%%%%%%%%%%%%%%%%%%%%%%%%%%%%%%%%%%%%%%%%%%%
Eckart's choice of $u^k$ along \R{6}$_1$
\bee{19}
u^k \ :=\ \frac{a^2}{n}N^k , \quad a\equiv c,\quad\mbox{or}\quad n^k\equiv 0,
\ee 
is more general than \R{17}: It does not restrict the energy-momentum tensor
or the spin tensor, because $N^k$ is not a part of $T^{ik}$ or 
$S_{ik}^{\cdot\cdot l}$. A second advantage is 
its illustrative interpretation: because 
the particle flux is purely convective and has no conductive part, 
$u^k$ is according to \R{19} the material 4-velocity, and we obtain
for the particle number flux according to Eckart
\bee{20}
N^k \ =\ \frac{1}{c^2}nu^k ,
\ee
an expression which is widely accepted in relativistic continuum physics.

\Section{Entropy Balance\label{EB}}

We now introduce Eckart's version into the entropy identity \R{13}
\bee{21}
S^k \ \equiv\ (s^k - \lambda q^k - \Lambda^m \Xi_m^{\cdot k}) 
+ (\mu N^k + \xi _l T^{kl} + \zeta^{nm}S_{nm}^{\cdot\cdot k}),
\ee
and \R{14} and \R{14a} are still valid. 
\vspace{.3cm}\newline%%%%%%%%%%%%%%%%%%%%%%%%%%%%%%%%%%%%%%%%%
In order to determine the entropy vector in accordance with this identity, 
one can exploit the entropy balance \R{2} 
\bee{2aa}
S^k\se{k}\ =\ \f + \sigma,
\ee
and by differentiating \R{21} and by use of the balance equations
\R{1}, we obtain
\bey\nonumber
S^k\se{k}\ =\ (s^k - \lambda q^k - \Lambda^m \Xi_m^{\cdot k})\se{k} +
\hspace{3cm}\\ \label{22} 
+ \mu\ko{k}N^k +
\xi_l {\se{k}} T^{kl} + \zeta^{nm}\se{k} S_{nm}^{\cdot\cdot k} +
\xi _l [G^l + K^{l}] + \zeta^{nm}(H_{[nm]} + L_{[nm]}) .
\eey
To interpret this balance by physics, one has to identify the supply and 
production terms $\f$ and $\sigma$. To this end, we refer to classical 
thermodynamics which defines the entropy supply as the energy supply $r$ times
the reciprocal rest temperature
\bee{23b1}
\f\ :=\ \frac{r}{T}.
\ee
The energy supply itself is caused by the external forces $K^i$ and by the 
external moments $L_{[ik]}$. Consequently, we have by definition
\bee{b23}
r\ :=\ u_i K^i + s_{lm}\Theta^{[lm][ik]}L_{[ik]}.
\ee
The tensor $\Theta^{[lm][ik]}$ which connects the spin to the external moments 
does not
need to be specified for our purposes. Interesting is that the rest
temperature $T$ is introduced by $T=r/\f$ according to \R{23b1}.  
\vspace{.3cm}\newline%%%%%%%%%%%%%%%%%%%%%%%%%%%%%%%%%%%%%%%%%
The entropy supply can be read off from \R{22}, and a comparison with \R{23b1}
results in
\bee{c23}
\f\ =\ \xi _l K^{l} + \zeta^{nm}L_{[nm]}\
=\ \frac{1}{T}u_i K^i + \frac{1}{T}s_{lm}\Theta^{[lm][ik]}L_{[ik]}.
\ee
This enables one to determine $\lambda$ and $\Lambda^m$ which were 
arbitrary up to now. From \R{14a}$_2$ and \R{14a}$_3$ follows
\byy{d23}
\xi _i  &=& \frac{u_i}{T}\ =\ \lambda u_i \\ \label{e23}
\zeta^{[ik]}&=& \frac{1}{T}s_{lm}\Theta^{[lm][ik]}\ =\ 2u^{[i}h^{k]}_m 
\Lambda^m \ =\ u^{i}h^{k}_m \Lambda^m - u^{k}h^{i}_m \Lambda^m .
\eey
Multiplication of \R{e23} with $u_i$ and taking \R{11c} into consideration 
results in
\bee{e23a}
\lambda\ =\ \frac{1}{T}, \qquad \Lambda^k \ =\ \frac{1}{c^2}
\frac{s_{lm}u_i}{T}\Theta^{[lm][ik]} .
\ee
The vector \R{d23} which is in accordance with the former definition 
\R{b11}$_3$ is called the 4-temperature. The vector \R{e23a}$_2$
which later on will play a role for formulating the
equilibrium conditions of the spin is called the temperature-spin.
\vspace{.3cm}\newline%%%%%%%%%%%%%%%%%%%%%%%%%%%%%%%%%%%%%%%%%%%%
After having determined the supply terms according to \R{d23} and \R{e23},
the remaining terms on the 
left-hand side of \R{22} have to be considered as the entropy production
according to \R{2aa}
\bey\nonumber
\sigma\ =\
(s^k - \frac{1}{T} q^k - \Lambda^m \Xi_m^{\cdot k})\se{k}\ +
\frac{1}{T}u_l G^l + \frac{2}{T}u^{[i}\Lambda^{k]} H_{[ik]}+
\hspace{.7cm}
\\ \label{f23}
+\ \mu\ko{k}N^k +
(\frac{1}{T}u_l ){\se{k}} T^{kl} + 2(u^{[n}\Lambda^{m]})\se{k} 
S_{nm}^{\cdot\cdot k}\ \geq\ 0.
\eey
This expression includes three terms of different characters,
a divergence term of a space-like vector, the SMEC-terms and 
terms according to the usual 
flux-force scheme \C{GRMA61} of the entropy production.
The divergence term contains fluxes which does not contribute to the
entropy production. Therefore we define the entropy flux by
\bee{g23} 
s^k \ :=\ \frac{1}{T} q^k + \Lambda^m \Xi_m^{\cdot k}.
\ee
Finally taking \R{g23} into account, the entropy production \R{f23} results in
\bee{h23} 
\sigma\ =\ \frac{1}{T}u_l G^l + \frac{2}{T}u^{[i}\Lambda^{k]} H_{[ik]}+
\mu\ko{k}N^k +
(\frac{1}{T}u_l ){\se{k}} T^{kl} + 2(u^{[n} \Lambda^{m]} )\se{k} 
S_{nm}^{\cdot\cdot k}\ \geq\ 0.\vspace{.3cm}
\ee%%%%%%%%%%%%%%%%%%%%%%%%%%%%%%%%%%%%%%%%%%%%%%%%%%%
The entropy follows from \R{21}, \R{d23} and \R{e23} 
\bee{i23}
S^k \ =\ \mu N^k + \frac{1}{T}u_l T^{kl} + 2u^{[n} \Lambda^{m]} 
S_{nm}^{\cdot\cdot k}.
\ee
In contrast to the entropy production, the entropy does not contain
SMEC-terms which are generated by differentiation.
For vanishing spin density, \R{i23} coincides with the ansatz 
\R{8d}. But \R{i23} is a derived relation and not only a guessed
ansatz. Beyond that, it includes
the spin, and also the entropy flux density \R{g23} and the entropy production
density \R{h23} follow consistently by the same procedure including the spin.
\vspace{.3cm}\newline
In \C{NEU80} the possibility is
briefly discussed, if the ansatz \R{8d} for the entropy can be
extended by adding a time-like vector. This possibility is excluded
by the identity \R{21} which allows to add only a space-like vector, the
first bracket in \R{21}.
\vspace{.3cm}\newline%%%%%%%%%%%%%%%%%%%%%%%%%%%%%%%%%%%%%%%%%
The entropy \R{i23} has the form of a sum of products which can be
written symbolically
\bee{j23}
S^k \ =\ {\bf X}\circ {\bf Y}^k .
\ee
The entropy balance equation \R{2aa} becomes
\bee{k23}
S^k\se{k} \ =\ {\bf X} \se{k}\circ {\bf Y}^k + {\bf X}\circ 
{\bf Y}^k\se{k}\ =\ \sigma + \f . 
\ee
According to \R{h23}, we obtain only for space-times of vanishing
SMEC-terms, $G^l \equiv 0,\ H_{[ik]}\equiv 0$
\bee{l23}
\sigma\ =\ {\bf X} \se{k}\circ {\bf Y}^k \quad\rightarrow\quad
\f \ =\ {\bf X}\circ {\bf Y}^k\se{k}.
\ee
According to \R{h23}, the entropy production density has not the usual 
form of a product of ``forces'' ${\bf X} \se{k}$ and ``fluxes'' ${\bf Y}^k$,
because the SMEC-terms vanish only for special space-times, but
not in general.

\Section{Equilibrium Conditions\label{EC}}

Equilibrium states are defined by equilibrium conditions. We have to 
distinguish between necessary and supplementary equilibrium conditions.
The necessary and the supplementary equilibrium conditions together represent
sufficient equilibrium. We will mark both kinds of 
equilibrium conditions differently: the necessary ones by $\seq$, the
supplementary ones by $\doteq$. For the present, we consider the
necessary conditions in the next section.

\subsection{Necessary equilibrium conditions}

The necessary equilibrium conditions are given by vanishing entropy
production density \R{h23} and vanishing entropy supply density \R{c23}
\bee{30}
\sigma\eq\ \seq\ 0, \quad \ \f\eq\ \seq\ 0\quad \longrightarrow\quad
S^k\se{k}\eqo\ =\ 0.
\ee
(equilibrium quantities are marked by $\eq$ or by $\eqo$ in the sequel) and
vanishing entropy flux density
\bee{a30}
s^k \eq\ \seq\ 0.
\ee
The implication in \R{30} follows from \R{2aa}.
\vspace{.3cm}\newline%%%%%%%%%%%%%%%%%%%%%%%%%%%%%%%%%%%%%
For the present, we will exploit the entropy supply density \R{30}$_2$ 
by starting out with \R{c23}.
Because the force $K^i$ is independent of the 
momentum $L_{[ik]}$, the part of the necessary equilibrium conditions 
belonging to the entropy supply splits into two parts and using
\R{e23}and \R{11c}, we obtain
\bee{30a1}
u_i\eqo K^i\eq \ =\ 0,\qquad
2(u^{[i}\Lambda^{k]})\eq L_{[ik]}\eqo\ =\ 0 .
\ee
From \R{30a1} we read off, that for the present neither the external 
forces nor the external moments have to be zero in equilibrium. Using
the balance equations \R{1}$_{2,3}$, we obtain
\bee{30a2} 
u_i\eqo[T^{ki}\se{k}-G^i]\eq\ =\ 0,\qquad
(u^{[i}\Lambda^{k]})\eq [S_{ik}^{\cdot\cdot j}{\se{j}} - H_{[ik]}]\eq\ = 0.
\vspace{.3cm}\ee%%%%%%%%%%%%%%%%%%%%%%%%%%%%%%%%%%
From \R{i23} follows by \R{30}$_3$
\bee{30a3}
0\ =\ \left(\mu N^k \right)^{eq}\se{k} + 
\left(\frac{1}{T}u_l T^{kl}\right)^{eq}\se{k} + 
2\left(u^{[n} \Lambda^{m]}S_{nm}^{\cdot\cdot k}\right)^{eq}\se{k}.
\ee
The $N^k$, $T^{kl}$ and $S_{nm}^{\cdot\cdot k}$ are not 
independent of each other, because they are coupled by constitutive
equations and by the SMEC-terms. Therefore we cannot state that 
each term of 
the sum \R{30a3} vanishes. The equilibrium condition \R{30a3}
is only one equation which cannot describe equilibrium completely.
Therefore we need supplementary equilibrium 
conditions beyond \R{30}and \R{a30}. These conditions will be considered in the
next section.

\subsection{Supplementary equilibrium conditions}

\subsubsection{Supply conditions}

According to the necessary condition \R{30a1}$_1$, the power of the external 
forces is zero in equilibrium. From that one cannot conclude that the
external forces vanish themselves in equilibrium. There exist an easy
criterion for testing whether the external forces vanish in equilibrium: 
Starting out again with \R{30a1}$_1$, we see that in equilibrium the
4-component of the force is zero in the rest system, marked by $^R$,
\bee{a30a3}
^R K^4\eq \ =\ 0.
\ee
If now also the 3-components of the force vanish in the rest system
\bee{1a30a3}
^R K^\alpha\eq \ =\ 0,\qquad\alpha=1,2,3,
\ee
we obtain the very special supplementary equilibrium condition
\bee{b30a3}
K^i\eq\ \doteq\ 0
\ee
for the external forces.
\vspace{.3cm}\newline%%%%%%%%%%%%%%%%%%%%%%%%%%%%%%%%%%%%%%%%%%%%%%%%%%%
According to \R{e23a}$_2$, the necessary equilibrium condition \R{30a1}$_2$,
depends on the spin density \R{10a}$_1$ and the temperature.
There may be non-zero $\Lambda^k \eq$-fields depending on the external 
moments in such a way that \R{30a1}$_2$ is satisfied,
but this situation is so strange, that we do not take this seriously into 
consideration. Consequently, we obtain two supplementary equilibrium conditions
\bee{37}
\Lambda^k\eq\ \doteq\ 0\qquad\cup\qquad L_{[ik]}\eqo\ \doteq\ 0,
\ee
that means, the external moments have to vanish in equilibrium in systems 
of non-vanishing spin. If the external moments do not vanish, the system must
be spin-free in equilibrium. These statements are true except for the exotic
situation that \R{30a1}$_2$ is satisfied for non-vanishing $\Lambda^k\eq$
and $L_{[ik]}\eqo$.

\subsubsection{N$^k$-Condition}

To begin with the supplementary equilibrium conditions, we consider by
use of \R{1}$_1$, \R{20} and the abbreviation $\td :=\ \se{k}u^k$
\bee{30a4}
\left(\mu N^k \right)\se{k}\ =\ \mu\ko{k} \frac{1}{c^2}nu^k \ =\
\frac{1}{c^2}n \mu\td .
\ee
Because it is obvious that there are no non-vanishing material time 
derivatives 
in equilibrium except that of the acceleration $u^k \td$
\footnote{The $^{\bullet}$ is the relativistic analogue to the 
non-relativistic material time derivative $d/dt$ which describes the time rates
of a rest-observer. Therefore, $d/dt$ is observer-independent and zero in
equilibrium \C{MURE02,MURE08}.}, 
we demand as a first supplementary equilibrium condition
\bee{30a5}
\boxplus\td\ \doteq\ 0,\quad \boxplus\ \neq\ u^k
\qquad\longrightarrow\qquad\left(\mu N^k \right)\se{k}\eqo\ =\ 0  .
\ee 
Consequently, we obtain by \R{20}
\bee{30a6}
0\ =\ \left(\mu N^k \right)\se{k}\eqo\ =\ \left(\mu \frac{1}{c^2}nu^k
\right)^{eq}\se{k}\ =\ \left(\mu \frac{1}{c^2}n\right) ^{\td}_{eq} + \mu
\frac{1}{c^2}n u^k\se{k}\eqo 
\ee
which by \R{30a5}$_1$ results in
\bee{30a7}
u^k\se{k}\eqo \ =\ 0. 
\ee
Further we obtain by \R{30a7} and \R{30a5}$_1$
\bee{30e2}
\left(\frac{1}{T}u^k \right)\!\se{k}\ =\ \frac{1}{T}u^k \se{k} + 
\left(\frac{1}{T}\right)\!^{\td}\quad\longrightarrow\quad
\left(\frac{u^k}{T}\right)\!\se{k}\!^{\eqo}\ =\ 0 .
\ee
Hence, the vanishing first term in \R{30a3} is exploited by applying the
supplementary equilibrium condition \R{30a5}. Now we will consider the
next term.

\subsubsection{T$^{kl}$-Condition}

Using \R{11}$_1$, we obtain for the second term in \R{30a3}
\bee{30e3} 
\left(\frac{u_l}{T}T^{kl}\right)\!\se{k}\ =\ 
\left(\frac{q^k}{T}+\frac{1}{c^2 T}eu^k \right)\!\se{k}\ =\
\left(\frac{q^k}{T}\right)\se{k} + \left(\frac{e}{c^2 T}\right)^{\td} +
\frac{e}{c^2 T}u^k\se{k} 
\ee
which by use of \R{30a5}$_1$ and \R{30a7} results in
\bee{30e4}
\left(\frac{u_l}{T}T^{kl}\right)^{eq}_{\se{k}}\ =\
\left(\frac{q^k}{T}\right)^{eq}\se{k}\ =\
\left(\frac{1}{T}\right)^{eq}_{\ko{k}}q^k_{eq} + 
\left(\frac{1}{T}\right)^{eq}q^k\se{k}{^{eq}}.
\ee
The first term of the right-hand side represents the 
dissipation due to heat conduction which is zero in equilibrium, a statement
which represents an other supplementary equilibrium condition
\bee{e30d2}
\left(\frac{1}{T}\right)_{\ko{k}}^{eq}q^k_{eq}\ \doteq\ 0\quad
\longrightarrow\quad
q^k \eq\ =\ 0.\vspace{.3cm}
\ee%%%%%%%%%%%%%%%%%%%%%%%%%%%%%%%%%%%%%%%%%%%%%%%%%%
Because there are equilibria with non-vanishing temperature gradient
(e.g. in gravitational fields) and because the dissipation due to heat
conduction is always not negative 
\bee{e30d3}
\left(\frac{1}{T}\right)_{\ko{k}} q^k\ \geq\ 0,
\ee
and the heat flux density depends continuously on the temperature
gradient, the conclusion in \R{e30d2} is the only possible one
\C{MUPAEH01}. 
But it is also obvious that there are no heat fluxes in equilibrium.
From \R{e30d2}$_2$ follows
\bee{d3e30}
q^k\se{k}\eqo\ =\ 0,\qquad \left(\frac{q^k}{T}\right)^{eq}\ =\ 0.
\vspace{.3cm}\ee%%%%%%%%%%%%%%%%%%%%%%%%%%%%%%%%%%%%%%%%%%%%%%%%%%
Taking \R{e30d2}$_2$ into account, \R{30e4} results by use of \R{30a2}$_1$ in
\bee{e30d4}
0\ =\ \left(\frac{u_l}{T}T^{kl}\right)^{eq}_{\se{k}}\ =\
\left(\frac{u_l}{T}\right)^{eq}\se{k}T^{kl}_{eq} + \frac{u_l}{T}^{eq}G^l_{eq}
\vspace{.3cm}\ee%%%%%%%%%%%%%%%%%%%%%%%%%%%%%%%%%%%%%%%%
We now consider the special case that the energy-momentum
tensor is symmetric in equilibrium (what is not the case in general). 
Then \R{e30d4} results in
\bee{30k}
T^{[kl]}_{eq}\ \equiv\ 0\quad\rightarrow\quad 
(\frac{1}{T}u_{(l} )_{;k)}\eqo T^{kl}\eq\ =\ -\frac{u_l}{T}^{eq}G^l_{eq}.
\ee
In general, we cannot conclude from \R{30k}$_2$, that the 
temperature 4-vector $u_l /T$ is killing in equilibrium even for SMEC-free 
space-times, 
\bee{30l} 
(\frac{1}{T}u_{l} )_{\se{k}}\eqo + (\frac{1}{T}u_{k} )_{\se{l}}\eqo\ 
\stackrel{?}{=}\ 0,
\ee
because we do not presuppose a
symmetric $T^{kl}$, as it was assumed in \C{NEU80}. Presupposing \R{30l},
no additional equilibrium conditions would follow for the symmetric
$T^{kl}$, because \R{30k} is satisfied identically for SMEC-free
space-times. We now treat the general case. 
\vspace{.3cm}\newline%%%%%%%%%%%%%%%%%%%%%%%%%%%%%%%%%%%%%%%%
After a short calculation, we obtain in non-equilibrium by using \R{7} 
\bee{e30a}
\left(\frac{u_l}{T} \right)\!{\se{k}} T^{kl}\ =\
\frac{1}{T}u_l\td\frac{1}{c^2}p^l + \frac{1}{T}u_{l}{\se{k}}t^{kl} +
\left(\frac{1}{T}\right)^\td\frac{1}{c^2}e + 
\left(\frac{1}{T}\right)_{\ko{k}}q^k .
\ee
Inserting
\bee{1e30a}
u_l \td p^l\ =\ -u_l p^l \td 
\ee
and taking \R{30a5}$_1$ and \R{e30d2}$_2$ into account, we obtain in
equilibrium
\bee{2e30a}
\left(\frac{u_l}{T} \right)\!{\se{k}}^{eq} T^{kl}\eq\ =\
\frac{1}{T} u_l {\se{k}}^{eq} t^{kl}\eq ,
\ee
and \R{e30d4} results in
\bee{3e30a}
0\ =\
u_l {\se{k}}^{eq} t^{kl}\eq  + u_l^{eq}G^l_{eq}.\vspace{.3cm}
\ee%%%%%%%%%%%%%%%%%%%%%%%%%%%%%%%%%%%%%%%%%%%%%%%%%%%%%
As we can see easily, the following identity is valid
\bee{3e30a1}
0\ =\
u_l {\se{k}}^{eq} t^{kl}\eq  + u_l^{eq}G^l_{eq}\ =\
u_l {\se{k}}^{eq}\left[t^{kl}\eq + \frac{u_p^{eq}G^p_{eq}}
{u_p{\se{q}}^{eq}A^{qp}}A^{kl}\right],
\ee
for all $A^{kl}$ with
\bee{3e30a2}
u_p{\se{q}}^{eq}A^{qp}\ \neq\ 0,\qquad u_p\td\eq u_q\eqo A^{qp}\ =\ 0.
\ee
We need the second property for later use.
Consequently, we can introduce non-unique modified stress tensors
which include the SMEC-term
\bee{3e30a3}\tau^{kl}\ :=\ t^{kl}\eq + J^{kl},\qquad
J^{kl}\ :=\ \frac{u_p^{eq}G^p_{eq}}{u_p{\se{q}}^{eq}A^{qp}}A^{kl}
\ee
and \R{3e30a} results in
\bee{9e30a}
0\ =\ u_l {\se{k}}^{eq} \tau^{kl}, 
\ee
a result which can be also expressed in an other way.
\vspace{.3cm}\newline%%%%%%%%%%%%%%%%%%%%%%%%%%%%%%%%%%%%%
As usual \C{NEU80}, we introduce the kinematical invariants by the
following definitions
\byy{31}
\mbox{shear:}\hspace{2cm}\sigma_{ab} &:=& u_{(a}{\se{b}}{_)} 
- \frac{1}{c^2}u\td _{(a}u_{b)} -
\frac{1}{3}\Theta h_{ab},\\ \label{31a}
\mbox{rotation:}\hspace{2cm}\omega_{ab} &:=& u_{[a}{\se{b}}{_]} 
- \frac{1}{c^2}u\td_{[a}u_{b]},\\ \label{31b}
\mbox{acceleration:}\hspace{2cm}u\td_a &:=& u_{a}{\se{b}}u^b ,\\ \label{31c}
\mbox{expansion:}\hspace{2.2cm}\Theta &:=& u^a \se{a}.
\eey
According to their definitions, we obtain from \R{5} and from the
normalization of the 4-velocity \R{4} and \R{19}$_2$
\bee{31d} 
h_{ab}u^b\ =\ u\td _a u^a \ =\ \sigma_{ab}u^b \ =\ \omega_{ab}u^b
\ =\ 0.
\ee
From \R{31} and \R{31a} follows the velocity gradient
\bee{31e}
u_{a}{\se{b}}\ =\ \sigma_{ab} + \omega_{ab} + \frac{1}{3}\Theta h_{ab} +
\frac{1}{c^2}u\td_a u_b ,
\ee
and we obtain
\bee{31e1}
u_{l}{\se{k}}\tau^{kl}\ =\ [\sigma_{lk} + \omega_{lk}]\tau^{kl} + 
\frac{1}{3}\Theta h_{lk}\tau^{kl} + \frac{1}{c^2}u\td_l u_k \tau^{kl}. 
\ee
By use of \R{30a7} and \R{3e30a2}$_2$, \R{9e30a} results in
\bee{a31e}
u_{l}{\se{k}}\eqo \tau^{kl}\ =\
[\sigma_{lk} + \omega_{lk}]^{eq}\tau^{kl}\ =\ 
\sigma_{lk} ^{eq}\tau^{(kl)} +
\omega_{lk}^{eq}\tau^{[kl]}\ =\ 0. 
\ee
Because the symmetric and the antisymmetric part of the stress tensor are
independent of each other, we can split \R{a31e} into
\bee{d31e}
\sigma_{lk}^{eq}\tau^{(kl)}\ =\ 0,\qquad
\omega_{lk}^{eq}\tau^{[kl]}\ =\ 0.
\ee
Because the tensor $A^{kl}$ in \R{3e30a3}$_2$, and consequently also $J^{kl}$ 
in \R{3e30a3}$_1$, can be chosen arbitrarily,  
the SMEC-term can be distributed freely on the
shear or rotation terms: If $A^{kl}$ is chosen to be symmetric, no
part of the SMEC-term appears in the rotation part and vice-versa.  
\vspace{.3cm}\newline%%%%%%%%%%%%%%%%%%%%%%%%%%%%%%%%%%%%%%%%%
The equilibrium conditions \R{d31e} can be interpreted differently: 
If we are looking
for equilibrium conditions which are the same for all space-times and 
materials, that means, they are valid for arbitrary $\tau^{(kl)}$ and 
$\tau^{[kl]}$, we obtain
\bee{e31e}
\sigma_{lk}^{eq} \ \doteq\ 0,\qquad\omega_{lk}^{eq} \ \doteq\ 0
\ee
as supplementary equilibrium conditions.
\vspace{.3cm}\newline%%%%%%%%%%%%%%%%%%%%%%%%%%%%%%%%%%%%%%%%%%%%%%%%%%
The second interpretation is as follows: Because \R{d31e} are derived
material-independently, there may be shear and rotation fields different
from zero satisfying \R{d31e} for special chosen space-times and
materials. That means,
there are special material- and space-time-dependent equilibria having 
non-vanishing shear and/or
rotation. By these remarks, the second necessary equilibrium condition 
\R{e30d4} is exploited, and we have now to consider the equilibrium
conditions belonging to the spin.

\subsubsection{S$^{\cdot\cdot k}_{nm}$-Condition}

Taking \R{d3e30}$_2$ and the necessary equilibrium condition \R{a30}
into account, we obtain from \R{g23} and \R{30a5}$_1$
\bee{d4e30} 
\Lambda^m_{eq}\Xi^{\cdot k}_m \eqo\ =\ 0.
\ee
Because the spin stress $\Xi^{\cdot k}_m$ is not regular 
\bee{d5e30}
\Xi^{\cdot k}_m u_k\ =\ 0,\qquad u^m \Xi^{\cdot k}_m\ =\ 0,
\ee
according to \R{10b}$_2$, and
\bee{d5e30a}
\Xi^{\cdot k}_m \eqo u_k \td\eq\ =\ 0,\qquad 
u^m \td\eq\Xi^{\cdot k}_m \eqo\ =\ 0,
\ee
according to \R{30a5}$_1$ and \R{d5e30}, there is the possibility that in 
equilibrium non-zero temperature-spins are
in the kernel of the spin stress as solution of \R{d4e30}.  
\vspace{.3cm}\newline%%%%%%%%%%%%%%%%%%%%%%%%%%%%%%%%%%%%%
We obtain from \R{30a3} by \R{30a5}$_2$ and \R{e30d4} an other necessary
equilibrium condition  
\bee{e30d5}
\left(u^{[n} \Lambda^{m]}S_{nm}^{\cdot\cdot k}\right)^{eq}\se{k}\ =\ 0.
\ee
As derived in appendix 2,
\bee{32a}
u^n \Lambda^m S^{\cdot\cdot k}_{nm}\ =\ \frac{1}{2}\Lambda^m
\left(\Xi^{\cdot k}_m +\frac{1}{c^2}\Xi_mu^k\right)
\ee
is valid. Consequently, by taking \R{d4e30}, \R{30a7} and
\R{30a5}$_1$ into account, \R{e30d5} results in
\bee{32a1}
0\ =\ \left[\Lambda^m \Xi_mu^k\right]\se{k}^{eq}\ =\
(\Lambda^m \Xi_m)\td\eq\ 
%=\ \Lambda^m_{eq}\td\Xi_m \eqo ,
\ee
and according to \R{e23a}$_2$, we obtain
\bee{32a1a}
\Lambda^k \Xi_k \ =\ \frac{1}{Tc^2}s_{lm}\Theta^{[lm][ik]}u_i \Xi_k .
\vspace{.3cm}\ee%%%%%%%%%%%%%%%%%%%%%%%%%%%%%%%%%%%%%%%%%%%%%%%%%%%%%
The spin variables \R{10a}, that are the spin density $s_{nm}$ and the spin 
vector
$\Xi_m$, and the constitutive equations \R{10b}, that are the couple stress
$s_{nm}^{\cdot\cdot k}$ and the spin stress $\Xi_m^k$, are related
by the spin axioms \C{HERMU04}
\byy{C4}
s_{nm} &=& \frac{1}{2}\eta_{nmpq}u^p\Xi^q ,\\ \label{C5}
s_{nm}^{\cdot\cdot k} &=& \frac{1}{2}\eta_{nmpq}u^p\Xi^{qk}. 
\eey
Here $\eta$ is the Levi-Civita symbol. The spin axioms are caused by the fact
that there are only three spin fields and only nine constitutive spin equations
\C{HERMU04}.
\vspace{.3cm}\newline%%%%%%%%%%%%%%%%%%%%%%%%%%%%%%%%%%%%%%%%%%%%%%%%%%
Inserting \R{C4} into \R{32a1a} results in
\bee{C4a}
\Lambda^j \Xi_j \ =\ \frac{1}{2Tc^2}\eta_{lm}^{\cdot\cdot pq}\Theta^{[lm][ik]}
u_p\Xi_q u_i\Xi_k ,
\ee
and by taking \R{30a5}$_1$ into account, \R{32a1} becomes
\bee{C4b}
0\ =\ \eta_{lm}^{\cdot\cdot pq}\Theta^{[lm][ik]}_{eq} \Xi_q \eqo \Xi_k \eqo
(u_p \td\eq u_i \eqo + u_p \eqo u_i \td\eq).
\ee
The case of a non-linear coupling tensor is also included, because
\bee{C4c}
\Theta^{[lm][ik]}_{eq}\td (\Xi_p , \Xi^q_p)\ =\ 0
\ee
is valid.
\vspace{.3cm}\newline%%%%%%%%%%%%%%%%%%%%%%%%%%%%%%%%%%%%%%%%%%
Because in \R{C4b} the antisymmetric parts of the quadratic forms in (q,k) 
and (p,i) do not contribute, we obtain
\bey\nonumber
[\eta_{lm}^{\cdot\cdot pq}\Theta^{[lm][ik]}_{eq} +
\eta_{lm}^{\cdot\cdot pk}\Theta^{[lm][iq]}_{eq} +
\eta_{lm}^{\cdot\cdot iq}\Theta^{[lm][pk]}_{eq} +
\eta_{lm}^{\cdot\cdot ik}\Theta^{[lm][pq]}_{eq}]\hspace{2cm}\\ \label{C4e}
\Xi_q \eqo \Xi_k \eqo
(u_p \td\eq u_i \eqo + u_p \eqo u_i \td\eq)\
\ =\ 0.
\eey
The tensor of 4th order in the square bracket has the following properties:
it is symmetric in (q,k) and in (p,i), and it has an empty kernel according to
the coupling property \R{b23}. In appendix 3 is proven that the only solutions
of \R{C4e} are
\bee{C4f}
u_p\td\eq\ \not=\ 0\qquad\longrightarrow\qquad
\Xi_q \eqo\ =\ 0\quad\longleftrightarrow\quad\Lambda^q_{eq}\ =\ 0,
\ee
or
\bee{C4fq}
\Xi_q \eqo\ \not=\ 0\qquad\longrightarrow\qquad u_p\td\eq\ =\ 0.
\ee
Thus, we proved 
\begin{quote} 
If the acceleration does not vanish in equilibrium, the system has to be 
spin-free, and if the system is not spin-free, the acceleration has to vanish
in equilibrium.\vspace{.3cm}
\end{quote}
%%%%%%%%%%%%%%%%%%%%%%%%%%%%%%%%%%%%%%%%%%%%%%%%%%%%%%%%%%%%%%%%%%%%%%
The equilibrium conditions \R{C4f} and \R{C4fq} have to be comparable with 
\R{d4e30} and \R{d5e30a}. This results in
\byy{C4fw}
u_p\td\eq\ \not=\ 0\qquad&\longrightarrow&\qquad u_p\td\eq\ 
\in\ \ker\Xi_m^{\cdot p}\eq\ \cap\ \Xi_p \eqo\ =\ 0\\ \label{C4fr}
\Xi_p \eqo\ \not=\ 0\qquad&\longrightarrow&\qquad\Xi_p \eqo\
\in\ \ker\Xi_m^{\cdot p}\eq\ \cap\ u_p\td\eq\ =\ 0.
\eey
We obtain from \R{C4e} to \R{C4fr} that
equilibrium is possible in the following cases
\byy{C4ft}
u_p\td\eq\ =\ 0\ \cap\ \Xi_q \eqo &=& 0,\\ \label{C4ft1}
u_p\td\eq\ \not=\ 0\ \cap\ \Xi_q \eqo &=& 0\ 
\cap\ u_p\td\eq\ \in\ \ker\Xi_m^{\cdot p}\eq\\ \label{C4ft2}
\Xi_q \eqo\ \not=\ 0\ \cap\ u_p\td\eq &=& 0\
\cap\ \Xi_p \eqo\ \in\ \ker\Xi_m^{\cdot p}\eq .
\eey
As \R{C4ft1} and \R{C4ft2} show, constitutive properties may prevent
equilibrium. Whereas in equilibrium the acceleration is always in the 
kernel of the spin stress according to \R{d5e30}, it depends of the 
material, if the spin density vector is an element of the kernel of the spin
stress in equilibrium. According to \R{C4ft}, the equilibrium is material 
independent only in spin-free materials with zero acceleration.
There are no equilibria with $u_p\td\eq\ \not=\ 0$ and $\Xi_q \eqo\ \not=\ 0$.

\Section{Recollection}

Because thermodynamics and relativity theory in their classical 
phenomenological versions have a common field of applications, one is 
challenged to look for a relativistic continuum thermodynamics. In analogy 
to non-relativistic continuum thermodynamics, generally one tries to found 
this theory on the balance equations of energy, momentum, spin and entropy. 
However, these balances alone do not provide a complete set of conditions to 
determine all thermodynamic quantities. Thus, like in non-relativistic 
thermodynamics \C{MUPAEH01}, one has to add constitutive equations describing 
the material. Therefore it is difficult (or even impossible) to 
formulate the constitutive equations generally. Insofar, one cannot expect 
to find a general and complete axiomatic approach to relativistic continuum
thermodynamics. The approach which is considered here contains only those 
conditions which can be formulated without any reference to special classes 
of materials. The aim of the paper is to present these conditions (especially
for equilibrium), well knowing the often overseen fact that finally 
constitutive equations formulated in a corresponding state space must be 
supplemented to obtain a closed system of partial differential equations 
describing a respective class of materials.
\vspace{.3cm}\newline%%%%%%%%%%%%%%%%%%%%%%%%%%%%%%%%%%%%%%%%%%%%%
According to this program, we start out with the balance equations \R{1} 
and \R{2} 
which are most general for the following reasons:
\begin{enumerate}%%%%%%%%%%%%%%%%%%%%%%%%%%%%%%%%%%%%%%%%%
\item The balances \R{1} and \R{2} are valid in Minkowski space-time 
as well as in curved space-times which are characterized by a connection 
defining a covariant derivative (i.e. they are true also in Riemann-, 
Riemann-Cartan-, and metric-affine space-times),
\item The balances \R{1} and \R{2} imply external inputs (the right-hand 
sides of \R{1} and \R{2}), the so-called 
supplies of energy-momentum, spin, and entropy, and they imply the internal
source terms caused by the spin-momentum-energy couplimg (SMEC) depending 
on the chosen space-time.
\item Entropy supply and entropy production are distiguished in the entropy 
balance. Therefore, in contrast to other approaches, the dissipation 
inequality takes the correct form \R{aa2} including the supplies.
\item The ansatz \R{i23} for the entropy density 4-vector motivated by the 
proved identity \R{13} is most general. It is  different from ansatzes in the 
literature, where the 4-vector of entropy is chosen in such a way that it 
reflects certain features of non-relativistic relations like the Gibbs-Duhem 
equation or the Carnot-Clausius relation.
\end{enumerate}%%%%%%%%%%%%%%%%%%%%%%%%%%%%%%%%%%%%%%%%%
In case of a definite theory of gravitation, the connection, and thus the 
space-time, is specified (e.g. to be Riemann, Einstein-Cartan or 
metric-affine) and the balance equations are completed by gravitational field 
equations formulated on the respective space-time. Further, as a consequence 
of the gravitational field equations and the differential identities valid 
in the respective type of space-time, the input and SMEC terms in \R{1} have 
to be 
specified, too. Therefore, a (material-independent) theory only based on 
1.-- 3. is necessarily incomplete in a multiple manner, namely for the 
missing specification of the state space, the supplies and due to the yet 
missing specification of the considered gravitational theory. 
In particular, the equilibrium conditions depend on the gravitational
equations, too. For instance, as was shown for GRT \C{10}, most equilibrium 
conditions adhoc introduced in \C{NEU80} result from Einstein's equations.
Otherwise, the advantage 
of such an approach considered here is, that it represents a comparatively 
general framework for possible relativistic continuum thermodynamics.

\Section{Conclusions}

After having proved the unrenouncable entropy identity \R{13}, for the present 
the most general relativistic expression for the entropy density 4-vector 
was derived. It contains three parts belonging to particle current, 
energy-mo\-men\-tum and spin. After that, arguments are given in favor of 
Eckart's ansatz of the particle flux density 4-vector being parallel to 
the 4-velocity of the material under consideration.
As a consequence, entropy supply and entropy production can be determined as 
expressions of relativistic invariant terms given by the balances 
\R{1} of energy-momentum and spin.
\vspace{.3cm}\newline%%%%%%%%%%%%%%%%%%%%%%%%%%%%%%%%%%%%%%%%%%
As a further implication of the entropy identity \R{13} and Eckart's ansatz, 
it can be shown, that the entropy density 4-vector, ad hoc introduced in 
\C{NEU80}, is the correct one (in case of the theory of general 
relativity) except the missing spin part, while the entropy expression given 
in \C{11} contradicts the entropy identity \R{13}. The latter follows from 
the fact that the expression correctly given in \C{NEU80} must not be 
supplemented by a time-like vector, as it was supposed in \C{NEU80} and done 
in \C{11}.
\vspace{.3cm}\newline%%%%%%%%%%%%%%%%%%%%%%%%%%%%%%%%%%%%%%%%%%%%%%%%%
After the more general considerations, the second part of the paper is devoted
to material-independent equilibria in relativistic thermodynamics. For the
present, equilibrium is defined by necessary equilibrium conditions: 
According to the second law, entropy supply, entropy production and entropy
4-flux vanish in equilibrium. From this demand, four equations 
(\R{30}$_1$, \R{a30} and \R{30a1}) follow.
\vspace{.3cm}\newline%%%%%%%%%%%%%%%%%%%%%%%%%%%%%%%%%%%%%%%%%%%%%%%%%
The above mentioned four necessary equilibrium conditions are not
sufficient for equilibrium. Consequently, we have to complete these necessary
equilibrium conditions by supplementary ones. These 
{\em supplementary equilibrium conditions} are
\begin{itemize}
\item The vanishing entropy supply results in
\begin{quote}
1: the power \R{30a1}$_1$ generated by the forces has to vanish in 
equilibrium. Sufficient for vanishing power is the supplementary equilibrium
condition that the forces themselves are zero in equilibrium \R{b30a3}
\bee{s0} 
u_i\eqo K^i \eq\ =\ 0\qquad\longleftarrow\qquad K^i \eq\ =\ 0.
\ee
2: if the material is not spin-free, the external moments have to vanish 
\R{37}. If they do not, the system has to be spin-free
\byy{s0a}
\Lambda^k\eq\ \not=\ 0\quad\longrightarrow\quad L_{[ik]}\eqo\ =\ 0, \\
\label{s0b}
\Lambda^k\eq\ =\ 0\quad\longleftarrow\quad L_{[ik]}\eqo\ \not=\ 0.
\eey
\end{quote}
\item Stemming from the entropy production generated by particle flux density, 
\begin{quote}
3: the material time derivatives 
have to vanish in equilibrium, except that of the 4-velocity
\bee{s1}
\boxplus_m \td\eqo\ :=\ \boxplus_m{\se{k}}\eqo u^k \eq\ =\ 0,\qquad 
\boxplus_m\eqo\ \neq u_m . 
\ee
4: the expansion \R{30a7} has to vanish in equilibrium
\bee{s1a}
u^k\se{k}\eqo\ =\ 0.
\ee
\end{quote}
\item Stemming from the entropy production generated by the energy-momentum 
tensor
\begin{quote}
5: the heat 4-flux density \R{e30d2}$_2$ and the entropy 3-flux density 
\R{d3e30}$_2$ have to vanish in equilibrium
\bee{s2}
q^k\eq \ =\ 0\qquad\longrightarrow\qquad \left(\frac{q^k}{T}\right)^{eq}\ 
=\ 0. 
\ee
6: independently of material and space time, shear and rotation \R{e31e}
have to be zero in equilibrium
\bee{s2a}
\sigma^{eq}_{lk}\ =\ 0,\qquad \omega^{eq}_{lk}\ =\ 0.
\ee
\end{quote}
\item Stemming from the entropy production generated by the spin tensor
\begin{quote}
7: equilibrium is possible in the following cases
\byy{xC4ft}
u_p\td\eq\ =\ 0\ \cap\ \Xi_q \eqo &=& 0,\\ \label{xC4ft1}
u_p\td\eq\ \not=\ 0\ \cap\ \Xi_q \eqo &=& 0\ 
\cap\ u_p\td\eq\ \in\ \ker\Xi_m^{\cdot p}\eq\\ \label{xC4ft2}
\Xi_q \eqo\ \not=\ 0\ \cap\ u_p\td\eq &=& 0\
\cap\ \Xi_p \eqo\ \in\ \ker\Xi_m^{\cdot p}\eq .
\eey
8: according to \R{s0b}, external moments need not be zero in equilibrium
\end{quote}
\end{itemize}

\Section{Appendices}
\subsection{Appendix 1}

Using the projector \R{5}, we obtain
\bee{A1}
N^k \ =\ N^l \delta ^k_l \ =\ N^l (h^k_l + \frac{1}{a^2}u^k u_l)\ =\
\frac{1}{a^2}nu^k + n^k .
\ee
The same procedure yields
\bey\nonumber
T^{ik}\ =\ T^{lm}\delta^i_l \delta^k_m \ =\ 
T^{lm}(h^i_l + \frac{1}{a^2}u^i u_l )(h^k_m + \frac{1}{a^2}u^k u_m)\ =
\\ \nonumber
=\ T^{lm} h^i_l h^k_m + \frac{1}{a^2}T^{lm}u^i u_l h^k_m + 
\frac{1}{a^2}T^{lm}h^i_l u^k u_m + \frac{1}{a^4}T^{lm}u^i u_l u^k u_m
\ =\\ \label{A2}
=\ t^{ik} + \frac{1}{a^2}u^i p^k + \frac{1}{a^2}q^i u^k +
\frac{1}{a^4}eu^i u^k .
\eey

\subsection{Appendix 2}

By use of \R{7a}, we obtain
\beo\nonumber
&&u^n \Lambda^m S^{\cdot\cdot k}_{nm}\ =\\ \nonumber
&&=\ u^n \Lambda^m
\left[ u^k \left(\frac{1}{c^2}s_{nm}+\frac{1}{2c^4}u_n\Xi_m -
\frac{1}{2c^4}u_m\Xi_n \right) + \right.\\ \nonumber
&&\left. + s_{nm}^{\cdot\cdot k}+ \frac{1}{2c^2}u_n\Xi_m^k-
\frac{1}{2c^2}u_m\Xi_n^k\right]\ = \\ \nonumber
&& = u^n \Lambda^m u^k \frac{1}{c^2}s_{nm} + u^n \Lambda^m u^k
\frac{1}{2c^4}u_n\Xi_m - u^n \Lambda^m u^k\frac{1}{2c^4}u_m\Xi_n + \\
\nonumber
&& + u^n \Lambda^m s_{nm}^{\cdot\cdot k} 
+ u^n \Lambda^m \frac{1}{2c^2}u_n\Xi_m^k
-u^n \Lambda^m\frac{1}{2c^2}u_m\Xi_n  .\label{B1}
\eey
According to \R{10a} and \R{10b}, the terms \#1, \#3, \#4 and \#6 are zero.
By use of \R{4}$_1$, we obtain
\bee{B1a}
u^n \Lambda^m S^{\cdot\cdot k}_{nm}\ =\ 
\frac{1}{2c^2}\Lambda^m \Xi_m u^k + \frac{1}{2}\Lambda^m \Xi^k_m 
\ee
which immediately results in \R{32a}.

\subsection{Appendix 3}

Because the coupling tensor is presupposed as regular, the solutions of 
\R{C4e} are
\byy{C4e1}
[\eta_{lm}^{\cdot\cdot pq}\Theta^{[lm][ik]}_{eq} +
\eta_{lm}^{\cdot\cdot pk}\Theta^{[lm][iq]}_{eq} +
\eta_{lm}^{\cdot\cdot iq}\Theta^{[lm][pk]}_{eq} +
\eta_{lm}^{\cdot\cdot ik}\Theta^{[lm][pq]}_{eq}] &=& 0,\\ 
\label{C4e2}\cup\hspace{9.13cm}
\Xi_q \eqo &=& 0,\\ 
\label{C4e3}\cup\hspace{2.6cm}
(u_p \td\eq u_i \eqo + u_p \eqo u_i \td\eq) = 0\quad\longrightarrow\quad
u_p \td\eq &=& 0.\hspace{.6cm}
\eey
These results are valid for an arbitrary SMEC-term $H_{[ik]}$. 
\vspace{.3cm}\newline%%%%%%%%%%%%%%%%%%%%%%%%%%%%%%%%%%%%%%
For the present, we investigate \R{C4e1} in more detail.
Multiplication of \R{C4e1} with arbitrary $A_{(qp)}$, and $B_{(qi)}$ results in
\byy{C4f1}
A_{(pq)}[
\eta_{lm}^{\cdot\cdot pk}\Theta^{[lm][iq]}_{eq} +
\eta_{lm}^{\cdot\cdot iq}\Theta^{[lm][pk]}_{eq}]\ =\ 0,\\ \label{C4f2}
B_{(qi)}[\eta_{lm}^{\cdot\cdot pq}\Theta^{[lm][ik]}_{eq} +
\eta_{lm}^{\cdot\cdot ik}\Theta^{[lm][pq]}_{eq}]\ =\ 0.
\eey
Consequently, we obtain
\byy{C4f3}
\eta_{lm}^{\cdot\cdot pk}\Theta^{[lm][iq]}_{eq}\ =\ 
-\eta_{lm}^{\cdot\cdot iq}\Theta^{[lm][pk]}_{eq},\\ \label{C4f4}
\eta_{lm}^{\cdot\cdot pq}\Theta^{[lm][ik]}_{eq}\ =\ 
-\eta_{lm}^{\cdot\cdot ik}\Theta^{[lm][pq]}_{eq}.\vspace{.3cm}
\eey%%%%%%%%%%%%%%%%%%%%%%%%%%%%%%%%%%%%%%%%%%%%%%%%%%%%%%%%%%%%%%%
For discussing \R{C4f3}, we have to distinguish six different cases
\byy{C4f5}
&\mbox{1: }&(l,m,p,k)\ \in\ \mbox{even}{\cal P}(1,2,3,4),\\ \label{C4f6}
&\mbox{2: }&(l,m,p,k)\ \in\ \mbox{odd}{\cal P}(1,2,3,4),\\ \label{C4f7}
&\mbox{3: }&(l,m,i,q)\ \in\ \mbox{even}{\cal P}(1,2,3,4),\\ \label{C4f8} 
&\mbox{4: }&(l,m,i,q)\ \in\ \mbox{odd}{\cal P}(1,2,3,4),\\ \label{C4f9}
&\mbox{5: }&(l,m,p,k)\ \notin\ {\cal P}(1,2,3,4),\\ \label{C4f10}
&\mbox{6: }&(l,m,i,q)\ \notin\ {\cal P}(1,2,3,4).
\eey
Here ${\cal P}(1,2,3,4)$ means a permutation of the elements (1,2,3,4) which
can be even or odd. All cases refer to \R{C4f3}.
\byy{C4g}
&1\cap 3:&\ \Theta^{[lm][iq]}_{eq}\ =\ - \Theta^{[lm][pk]}_{eq}\ =\ 0,
\\ \label{C4h}
&1\cap 4:&\ \Theta^{[lm][iq]}_{eq}\ =\ + \Theta^{[lm][pk]}_{eq}\ =\ 0,
\\ \label{C4i}
&2\cap 3:&\ -\Theta^{[lm][iq]}_{eq}\ =\ - \Theta^{[lm][pk]}_{eq}\ =\ 0,
\\ \label{C4j}
&2\cap 4:&\ -\Theta^{[lm][iq]}_{eq}\ =\ + \Theta^{[lm][pk]}_{eq}\ =\ 0,
\\ \label{C4k}
&1\cap 6:&\ \Theta^{[lm][iq]}_{eq}\ =\ 0,
\\ \label{C4l}
&2\cap 6:&\ - \Theta^{[lm][iq]}_{eq}\ =\ 0,
\\ \label{C4m}
&3\cap 5:&\ 0\ =\ - \Theta^{[lm][pk]}_{eq},
\\ \label{C4n}
&4\cap 5:&\ 0\ =\ + \Theta^{[lm][pk]}_{eq},
\\ \label{C4o}
&5\cap 6:&\ 0\ =\ 0.\vspace{.3cm}
\eey%%%%%%%%%%%%%%%%%%%%%%%%%%%%%%%%%%%%%%%%%%%%%%%%%%%%%%%%%%%%%
A comparison of \R{C4f5} to \R{C4f10} with \R{C4g} to \R{C4n} results in
\bee{C4p}
\Theta^{[lm][pk]}_{eq}\ =\ 0, \qquad \mbox{for all l, m, p, k}.
\ee
Because the couple tensor $\Theta^{[lm][pk]}_{eq}$ does not vanish identically,
we have to dismiss \R{C4e1}, and \R{C4e2} and \R{C4e3} remain as possible 
solutions. 
\vspace{.3cm}\newline%%%%%%%%%%%%%%%%%%%%%%%%%%%%%%%%%%%%%%%%%%
Consequently, we obtain 
\bee{C4fa}
u_p\td\eq\ \not=\ 0\qquad\longrightarrow\qquad
\Xi_q \eqo\ =\ 0\quad\longleftrightarrow\quad\Lambda^q_{eq}\ =\ 0,
\ee
or
\bee{C4fb}
\Xi_q \eqo\ \not=\ 0\qquad\longrightarrow\qquad u_p\td\eq\ =\ 0.
\ee

\noindent
\vspace{1cm}\newline
{\bf Acknowledgement} W.M. thanks R. Wulfert for his co-working in 
sect. 2 and 3 of a former unpublished version of this paper.
\newline 


\begin{thebibliography}{99}
\bibitem{4}
G.A. Maugin: On the covariant equations of the relativistic
electrodynamics of continua I, J. Math. Phys. 19 (1978) 1198
\bibitem{HHBCH} H.-H. v. Borzeszkowski, T. Chrobok: On special- and 
general-relativistic thermodynamics, Atti dell' Academia Peloritana del 
Pericolanti. - Classe di Scienze Fisiche, Matematiche e Natural, 
Bd. LXXXV, C1S0701002 – Suppl. 1, 2007
\bibitem{OBPI} Yu.N. Obukhov, O.B. Piskareva: Spinning fluid in general 
relativity, Class. Quantum Grav. 6 (1989) L15
\bibitem{EC}
C. Eckart: Thermodynamics of irreversible proceses III, Phys. Rev. 58
(1940) 919
\bibitem{KL}
G.A. Kluitenberg, S.R. de Groot, P. Mazur: Relativistic thermodynamics
of irreversible processes I, II, Physica (1953) 689, 1071
\bibitem{3}
J.A. Schouten: Ricci-Calculus, Springer, Berlin 1954
\bibitem{5}
T.W.B. Kibble: Lorentz invariance and the gravitational field,
J. Math. Phys. 2 (1961) 212
\bibitem{6}
D.W. Sciama: On the analogy between charge and spin in general
relativity theory, in: Recent Developments in General Relativity,
Pergamon Press, Oxford and PWN-Polish Scientific Publishers,
Warszawa, 1962, p. 415
\bibitem{7}
F.W. Hehl, J.D. McCrea, E.W. Mielke, Y. Ne'eman: Metric-affine
theory of gravity: Field equations, Noether identities, world spinors
and breaking of dilatation invariance, Physics Reports 258, No. 1,
\S 2, 1995 
\bibitem{MUPAEH96}
W. Muschik, C. Papenfuss, H. Ehrentraut: Concepts of Continuum Thermodynamics,
Kielce University of Technology 1996, ISBN 83-905132-7-7
\bibitem{NEU80}
G. Neugebauer: Relativistische Thermodynamik, Vieweg 1980, ISBN 3-528-06863-9
\bibitem{DPG90}
Reported by W. Muschik at Relativit\"atstage Berlin - Jena, 
June 20 - 22, 1990, TU Berlin, Verhandl. DPG (VI) 25 (1990) 314
\bibitem{HERMU04}
H. Herrmann, W. Muschik, G. R\"{u}ckner, H.-H. von Borzeszkowski: Spin
axioms in different geometries of relativistic continuum physics, Foundations
of Physics 34 (2004) 1005
\bibitem{11}
W. Israel, J.M. Stewart: Progress in relativistic thermodynamics and
electrodynamics of continuous media, in: A. Held (Ed), General
Relativity and Gravitation, Vol. 2, Plenum Press, New York, 1997, p. 491
\bibitem{12}
L.D. Landau, E.M. Lifshitz: Fluid Mechanics, Addison-Wesley, 
Reading/Massachusetts, 1959, p. 505
\bibitem{GRMA61}
S.R. de Groot, P. Mazur: Non-Equilibrium Thermodynamics, North-Holland,
Amsterdam, 1961, ch.III \S 2
\bibitem{MURE02}
W. Muschik, L. Restuccia, Changing the observer and moving materials in
continuum physics: Objectivity and material frame-indifference, Technische
Mechanik, 22 (2002) 152
\bibitem{MURE08}
W. Muschik, L. Restuccia, Systematic remarks on objectivity and 
frame-indifference: Liquid crystal theory as an example, Arch. Appl. Mech.,
DOI 10.1007/s00419-007-0193-2
\bibitem{MUPAEH01}
W. Muschik, C.Papenfuss, H. Ehrentraut, A sketch of continuum thermodynamics,
J. Non-Newtonian Fluid Mech. 96 (2001) 255, (proposition on p.279)
\bibitem{10}
T. Chrobok, H.-H. v. Borzeszkowski: Thermodynamical equilibrium and
spacetime geometry, Gen. Rel. Grav. 38 (2006) 397


\end{thebibliography}
\end{document}